\documentclass{l4dc2027}

\title[Kernel-Guided Learning of Lifted Sparse Polynomial Models]{Kernel-Guided Learning of Lifted Sparse Polynomial Models of Nonlinear Dynamical Systems}
\usepackage{times}
\usepackage{booktabs}
\usepackage{tikz}
\usepackage{bm}
\usetikzlibrary{positioning,arrows.meta,shapes.geometric,calc,patterns}
\usepackage{enumitem}
\usepackage{microtype}
\usepackage{threeparttable}
\newtheorem{assumption}{Assumption}
\newcommand{\R}{\mathbb{R}}
\newcommand{\NN}{\mathbb{N}}

\newcommand{\E}{\mathbb{E}}
\newcommand{\D}{\mathcal{D}}
\newcommand{\Dv}{\mathcal{D}_{\val}}
\newcommand{\Xc}{\mathcal{X}}
\newcommand{\Zc}{\mathcal{Z}}
\newcommand{\Kc}{\mathcal{K}}

\newcommand{\Gc}{\mathcal{G}}
\newcommand{\Mc}{\mathcal{M}}

\newcommand{\Sc}{\mathcal{S}}
\newcommand{\Nc}{\mathcal{N}}
\newcommand{\Oc}{\mathcal{O}}
\DeclareMathOperator{\diag}{diag}
\DeclareMathOperator{\rank}{rank}
\DeclareMathOperator{\range}{range}
\DeclareMathOperator{\dist}{dist}
\DeclareMathOperator{\spn}{span}
\DeclareMathOperator*{\argmin}{arg\,min}

\newcommand{\val}{\mathrm{v}}
\newcommand{\test}{\mathrm{t}}
\newcommand{\Nv}{N_{\val}}
\newcommand{\Nt}{N_{\test}}
\newcommand{\ngen}{n_g}
\newcommand{\nz}{n_z}
\newcommand{\ntil}{\tilde{n}}
\newcommand{\nbar}{\bar{n}}
\newcommand{\dks}{d_k^\star}
\newcommand{\dkh}{\hat{d}_k}
\newcommand{\epsb}{\bar{\epsilon}}
\newcommand{\lamb}{\bar{\lambda}}
\newcommand{\Tmax}{T_{\max}}
\newcommand{\Ttmax}{\tilde{T}_{\max}}
\newcommand{\Thmax}{\hat{T}_{\max}}
\newcommand{\x}{\mathbf{x}}
\newcommand{\xh}{\hat{\x}}
\newcommand{\y}{\mathbf{y}}
\newcommand{\z}{\mathbf{z}}
\newcommand{\zt}{\tilde{\z}}
\newcommand{\dz}{\dot{\z}}
\newcommand{\w}{\mathbf{w}}
\newcommand{\f}{\mathbf{f}}
\newcommand{\fh}{\hat{\f}}
\newcommand{\g}{\mathbf{g}}
\newcommand{\F}{\mathbf{F}}
\newcommand{\Fh}{\hat{\F}}
\renewcommand{\P}{\mathbf{P}}
\newcommand{\Ph}{\hat{\P}}
\newcommand{\cc}{\mathbf{c}}
\newcommand{\ch}{\hat{\cc}}
\newcommand{\zero}{\mathbf{0}}
\newcommand{\A}{\mathbf{A}}

\newcommand{\Y}{\mathbf{Y}}
\newcommand{\Z}{\mathbf{Z}}
\newcommand{\K}{\mathbf{K}}
\newcommand{\I}{\mathbf{I}}
\newcommand{\V}{\mathbf{V}}

\newcommand{\W}{\mathbf{W}}
\newcommand{\U}{\mathbf{U}}
\newcommand{\Q}{\mathbf{Q}}
\newcommand{\Smat}{\mathbf{S}}
\newcommand{\Dmat}{\mathbf{D}}
\newcommand{\bpsi}{\bm{\psi}}
\newcommand{\bvarphi}{\bm{\varphi}}

\newcommand{\bxi}{\bm{\xi}}
\newcommand{\balpha}{\bm{\alpha}}

\newcommand{\beps}{\bm{\epsilon}}
\newcommand{\bepst}{\tilde{\bm{\epsilon}}}
\newcommand{\bTheta}{\bm{\Theta}}
\newcommand{\bPhi}{\bm{\Phi}}
\newcommand{\bSigma}{\bm{\Sigma}}
\newcommand{\bLambda}{\bm{\Lambda}}
\newcounter{panel}

\author{%
\Name{Mohammad Itani} \Email{mohammad.itani@tuhh.de}\\
\addr Institute of Control Systems, Hamburg University of Technology, Hamburg, Germany \\
\Name{Timm Faulwasser} \Email{timm.faulwasser@ieee.org}\\
\addr Institute of Control Systems, Hamburg University of Technology, Hamburg, Germany
}
\begin{document}
\maketitle
\thispagestyle{empty}
\begin{abstract}%
    Polynomial models of dynamical systems enable constructive analysis via sum-of-squares programming. Yet, many physical systems involve non-polynomial nonlinearities which for certain classes of systems may be inferred from the underlying physics even when the exact dynamics are unknown. We show that fitting lifted polynomial models directly to lifted data is problematic, since the exact lifting constraints render the lifted derivatives linearly dependent. Instead, we learn the dynamics in the extended state, i.e., the state together with the known nonlinearities, and obtain the dynamics of the nonlinearities via the exact lifting relations. To avoid the combinatorially large libraries that sparse regression requires when the polynomial degrees are unknown, we propose KINDy, which selects the degree of each component by kernel ridge regression before applying sparse regression to the correspondingly reduced library. We prove that, for sufficiently small noise and regularization, provided the data satisfy a rank condition, the selected degree never underestimates the true degree. On the cart-pole system, KINDy recovers the true degrees at low noise levels and achieves speedups of up to 8.5 over sparse regression on the full library, while in almost all cases yielding models that are at least as sparse and have lower or equal test errors.
\end{abstract}
\section{Introduction}
Learning models of dynamical systems from data, known as system identification \citep{ljung1998system}, is essential whenever the underlying physics is partially or completely unknown. However, learned models differ significantly in their interpretability. Neural networks \citep{narendra1990identification, chen2018neural} and Gaussian processes \citep{rasmussen2003gaussian,kocijan2005dynamic}, for instance, act as black-box function approximators that reproduce the input-output behavior without providing an explicit representation of the underlying governing equations. Conversely, symbolic regression \citep{bongard2007automated} seeks to identify explicit mathematical expressions from data by searching over possible combinations of variables, numerical constants, and mathematical operators. However, searching over all possible combinations can become computationally prohibitive, particularly for high-dimensional systems. Sparse regression techniques, on the other hand, restrict the search space to a predefined candidate library and obtain parsimonious models by algorithms such as sequentially thresholded least squares, as introduced with the sparse identification of nonlinear dynamics (SINDy) \citep{brunton2016discovering}, or its ridge variant (STRidge) \citep{rudy2017data}, whose hyperparameters can be selected by information criteria \citep{mangan2017model}. This is well motivated, as many physical systems are governed by only a few dominant terms \citep{brunton2016discovering}.

When the candidate library consists solely of polynomial functions of the state, sparse regression yields sparse polynomial dynamical systems. Such representations are attractive because they enable constructive and tractable system analysis via sum-of-squares (SOS) programming \citep{parrilo2000structured}, such as Lyapunov function synthesis \citep{papachristodoulou2002construction} or barrier certificate construction \citep{prajna2004safety}. However, many physical systems, like mechanical systems, involve non-polynomial terms such as trigonometric functions. While sparse regression easily accommodates these expressions within the candidate library, the resulting models cannot be directly analyzed via standard SOS methods. To bridge this gap, lifting via coordinate extension converts the model into an equivalent polynomial form of higher dimension that is subject to algebraic constraints in the DAE sense, allowing SOS-based verification to be applied to a large class of nonlinear systems \citep{papachristodoulou2005analysis}. This is closely related to immersion into polynomial systems \citep{ohtsuka2005model}, to transformations appearing in nonlinear model order reduction \citep{gu2011qlmor, kramer2019nonlinear}, and to recasting in biological systems analysis \citep{savageau1987recasting}. Unlike Koopman operator embedding \citep{koopman1931hamiltonian, mauroy2020koopman} or Carleman linearization \citep{carleman1932application, kowalski1991nonlinear}, which seek linear representations that in general require infinite dimensions and thus truncation, polynomial lifting keeps the representation finite-dimensional and exact.

For many classes of systems, the non-polynomial terms, which we refer to as \emph{generators}, are known from the underlying physics. Examples include trigonometric functions in power systems \citep{kundur1994power} or mechanical systems, where rational functions also arise \citep{spong2020robot}. The lifting can then be constructed explicitly, as shown in the proof of Proposition~\ref{prop:existence}. This raises the question of how to learn models in lifted coordinates from data, which we address in this paper. The most direct approach is to fit each component of the lifted model to lifted data. The lifted outputs, however, are sample-wise linearly dependent, and dependent outputs must be removed before fitting, as otherwise the resulting estimates are meaningless \citep{box1973some}. Moreover, independently learned components need not respect the chain rule that links the dynamics of the generators to those of the state, so that trajectories of the learned model may leave the manifold of lifted states, which is the set on which the lifted model represents the original system. To avoid both issues, we propose to learn in the \emph{extended state}, consisting of the state together with the generators, and then to obtain the lifted model through the chain rule, which keeps this manifold invariant by construction (Proposition~\ref{prop:invariance}).

Learning in the extended state with sparse regression requires a library of monomials in the state and the generators, whose maximum degree must be chosen conservatively to cover the most complex component, since the degrees are unknown. As the number of monomials grows combinatorially with this degree and the dimension of the extended state, STRidge with model selection becomes increasingly expensive, since the ridge regressions within STRidge are repeated for many hyperparameter values on the grid. Kernel ridge regression (KRR) with a polynomial kernel, in contrast, implicitly represents all monomials up to a given degree at a dual cost that depends on the number of data points rather than on the number of monomials \citep{scholkopf2001learning}, but yields models that are dense in the monomial basis. We therefore propose kernel-guided SINDy (KINDy), which uses KRR to select the degree of each component and then applies STRidge to the correspondingly reduced library. Provided the data satisfy a rank condition, the selected degree never falls below the true degree for sufficiently small noise and regularization (Theorem \ref{thm:no_underestimation}), while an overestimated degree only increases the computational cost. In a cart-pole study, KRR selects the true degrees at low noise levels, while KINDy is up to 8.5 times faster than STRidge on the full library and in almost all settings yields models with on average at most as many monomials as well as lower or equal median test errors.

The remainder of this paper is organized as follows. Section~\ref{sec:problem} formulates the problem, Section~\ref{sec:liftvsextend} compares learning from lifted data with learning in the extended state, Section~\ref{sec:KINDy} presents KINDy, Section~\ref{sec:cart_pole} reports the cart-pole study, and Section~\ref{sec:conclusion} concludes. The appendices contain the proofs of all results and supplementary material for the numerical example.
\paragraph{Notation.}
For a vector $\x \in \R^n$, $x_{(k)}$ and $\x_{i:j}$ denote its $k$-th component and the vector of its components $i$ to $j$, respectively. For a vector-valued function $\f$, $f_k$ denotes its $k$-th component, and for a differentiable map $\bpsi$, $D\bpsi(\x)$ denotes its Jacobian at $\x$. For a matrix $\A$, $\A_{:,k}$, $\rank\A$, and $\range\A$ denote its $k$-th column, its rank, and its column space, respectively. Moreover, $\I_n$ denotes the $n \times n$ identity matrix, and $\diag(\cdot)$ a diagonal matrix. The symbols $\R[\x]$, $\R[\x]^n$, and $\R(\x)$ denote the ring of polynomials in $\x$ with real coefficients, the set of $n$-dimensional vectors of such polynomials, and the field of rational functions in $\x$, respectively. The expectation is denoted by $\E[\cdot]$. The cardinality of a finite set $\Sc$ is denoted by $|\Sc|$. Computational costs are stated in the big-O notation $\Oc(\cdot)$.
\section{Problem formulation}
\label{sec:problem}
Consider a dynamical system
\begin{equation}\label{eq:sys}
    \dot{\x} = \f(\x), \qquad \x \in \Xc,
\end{equation}
where $\Xc \subseteq \R^n$ is a non-empty, open, and connected set and $\f \colon \Xc \to \R^n$ is an unknown, non-polynomial ($\f \notin \R[\x]^n$), locally Lipschitz vector field. We are given a training data set $\D = \{(\x_i, \y_i)\}_{i=1}^{N}$ and a validation data set $\Dv = \{(\x_j^{\val}, \y_j^{\val})\}_{j=1}^{\Nv}$, where the state measurements $\x_i, \x_j^{\val}$ are assumed to be noise-free and the derivative measurements $\y_i, \y_j^{\val}$ are corrupted by i.i.d.\ zero-mean Gaussian noise, i.e.,
\begin{equation*}
    \y_i = \f(\x_i) + \beps_i, \quad
    \y_j^{\val} = \f(\x_j^{\val}) + \beps_j^{\val}, \quad
    \beps_i, \beps_j^{\val} \sim \Nc(\zero, \bSigma), \quad
    \bSigma = \diag(\sigma_1^2, \dots, \sigma_n^2).
\end{equation*}

The following assumption describes what is known about the structure of $\f$.
\begin{assumption}
    \label{assump:DA}
    There exists a known real-analytic function $\g \colon \Xc \to \R^{\ngen}$ whose components $g_1, \ldots, g_{n_g}$ are non-polynomial and differential-algebraic such that $\f(\x)\in\R[\x, \g(\x)]^n$. \hfill$\square$
\end{assumption}
A real-analytic function $\mathbf{g}$ is called \emph{differential-algebraic} if, for every $l = 1, \ldots, n$, each of its components $g_j$, $j = 1, \ldots, n_g$, satisfies a nontrivial polynomial equation in $\mathbf{x}$, $g_j$, and finitely many of its partial derivatives with respect to $x_{(l)}$ \citep{rubel1985differentially, ohtsuka2005model}. This class includes virtually all real-analytic functions that appear in the modeling of physical systems. We refer to the components of $\g$ as \emph{generators}. In general, $\g$ cannot be identified from data alone. However, for many physical systems, $\g$ can be inferred from the underlying physics, e.g., rotational kinematics naturally gives rise to trigonometric functions. It is worth noting that some generators may depend on some system parameters \citep{viknesh2026adam}. We now stack the state $\x$ and the generators $\g$ into the \emph{extended state} $\z = (\x, \g(\x)) \in \R^{\nz}$, where $\nz = n + \ngen$. It follows from Assumption~\ref{assump:DA} that every component $f_k$ of $\f$ satisfies $f_k(\x) = P_k(\z)$ for some polynomial $P_k$. This polynomial is generally not unique, since the generators may satisfy algebraic identities such as $\sin^2 x_{(l)} + \cos^2 x_{(l)} = 1$, and we therefore define the \emph{true degree} of $f_k$ as
\begin{equation}
    \label{eq:dk_star}
    \dks = \min\bigl\{ \deg P : P \in \R[\z],\ f_k(\x) = P(\x, \g(\x)) \text{ for all } \x \in \Xc \bigr\},  
\end{equation}
where $\deg P$ denotes the total degree of $P$, i.e., the largest sum of exponents among its monomials. Henceforth, $P_k$ denotes a polynomial of degree $\dks$ with $f_k(\x) = P_k(\z)$, and we collect these polynomials in $\P = (P_1, \ldots, P_n)$.
\paragraph{Polynomial lifting.} Although each $f_k$ is polynomial in $\z$, the dynamics of $\z$, 
\(
\dz = \big(\f(\x),D\g(\x)\,\f(\x)\big),
\)
are not necessarily polynomial in $\z$. A sufficient condition for the polynomiality of $\dz$ in $\z$ is that $\g$ is closed under differentiation, since, in this case,  $D\g(\x)$ is also polynomial in $\z$. If $\g$ is not closed under differentiation, then $\z$ must be augmented with additional functions of $\x$, which motivates the following definition.
\begin{definition}[Polynomial lifting]\label{def:lifting}
    A map 
    \(
    \Psi \colon \Xc' \to \R^{\ntil},
    \)
    where $\ntil > n$ and $\Xc' \subseteq \Xc$ is open and dense in $\Xc$, is called a polynomial lifting of \eqref{eq:sys} if it is of the form $\Psi(\x) = (\x, \bpsi(\x))$, where $\bpsi \colon \Xc' \to \R^{\nbar}$ is real-analytic on $\Xc'$ and $\nbar = \ntil - n$, and if there exists a polynomial vector field $\F \colon \R^{\ntil} \to \R^{\ntil}$ such that
    \(
        D \Psi(\x) \, \f(\x) = \F(\Psi(\x))
    \) for all $\x \in \Xc'$. \hfill $\square$
\end{definition}
By the chain rule and Definition~\ref{def:lifting}, the \emph{lifted state} $\Psi(\x)$ along a trajectory of \eqref{eq:sys} evolves according to
\begin{equation}
    \label{eq:chain_lifted}
    \frac{\mathrm{d}}{\mathrm{d}t} \Psi(\x(t)) = D\Psi(\x(t))\, \f(\x(t)) = \F(\Psi(\x(t))),
\end{equation}
which motivates the \emph{lifted (polynomial) system}
\begin{equation}\label{eq:tilde_z_dyn}
    \dot{\zt} = \F(\zt), \qquad \zt \in \R^{\ntil}.
\end{equation}
Since $\F$ is polynomial (hence locally Lipschitz), \eqref{eq:tilde_z_dyn} has a unique solution $\zt(\cdot;\zt_0)$ on a maximal time interval $[0, \Ttmax)$ for every $\zt_0 = \zt(0) \in \R^{\ntil}$. Now consider the solution $\x(\cdot;\x_0)$ of \eqref{eq:sys} with initial condition $\x_0 = \x(0) \in \Xc'$, and denote by $\Tmax \in (0,\infty]$ its maximal existence time in $\Xc'$. By \eqref{eq:chain_lifted}, the corresponding \emph{lifted curve} $\Psi(\x(\cdot;\x_0))$ is a solution of \eqref{eq:tilde_z_dyn} on $[0, \Tmax)$ that starts at $\Psi(\x_0)$, and hence, by maximality and uniqueness of solutions of \eqref{eq:tilde_z_dyn}, we obtain $\Ttmax \geq \Tmax$ and
\begin{equation}\label{eq:invariance}
  \zt(t;\Psi(\x_0)) = \Psi(\x(t;\x_0))
  \qquad \text{for all } t \in [0, \Tmax).
\end{equation}
Hence, the state trajectory $\x(\cdot;\x_0)$ of \eqref{eq:sys} may be obtained by projecting $\zt(\cdot; \Psi(\x_0))$ onto the first $n$ coordinates, i.e., $\x(t; \x_0) = [\I_n\ \zero]\, \zt(t; \Psi(\x_0))$ on $[0, \Tmax)$. Moreover, the set $\Zc = \Psi(\Xc')$ is the graph of the real-analytic function $\bpsi$ and hence defines an $n$-dimensional real-analytic embedded submanifold of $\R^{\ntil}$. Additionally, by \eqref{eq:invariance} it is forward invariant under \eqref{eq:tilde_z_dyn} on $[0, \Tmax)$. The next result guarantees the existence of a polynomial lifting $\Psi$ of \eqref{eq:sys} under Assumption \ref{assump:DA}.
\begin{proposition}[Existence of $\Psi$]
    \label{prop:existence}
    If Assumption~\ref{assump:DA} holds, then there exists a polynomial lifting
    $\Psi(\x) = (\x, \bpsi(\x))$ of~\eqref{eq:sys} such that the first $\ngen$ components of $\bpsi$ are the generators $\g$ and $D\bpsi(\x) = \Q(\Psi(\x))$
    for all $\x \in \Xc'$, where $\Q$ is some polynomial matrix. \hfill $\square$
\end{proposition}
The proof of Proposition~\ref{prop:existence}, along with all other proofs, is given in Appendix~\ref{app:proofs}. Since this proof is constructive, it also shows how $\Psi$ is obtained from $\g$. Note, however, that a polynomial lifting $\Psi$ is generally not unique.
\paragraph{Problem statement.}
Given the data $(\D, \Dv)$, learn \emph{efficiently} a \emph{sparse polynomial vector field} $\Fh$ approximating $\F$ in the lifted system~\eqref{eq:tilde_z_dyn}. A sparse model is interpretable \citep{tibshirani1996regression,hastie2015statistical}, less prone to fitting spurious terms to the measurement noise \citep{brunton2016discovering,mangan2017model}, and tends to generalize better when the true dynamics are sparse \citep{kaiser2018sparse,champion2019data}, while the polynomial form enables SOS-based analysis such as the synthesis of Lyapunov or barrier functions \citep{parrilo2000structured,papachristodoulou2002construction,prajna2004safety}. Efficiency is required since the true degrees $d_k^\star$ are unknown, so that sparse regression must use large, conservatively chosen libraries whose size drives the computational cost.
\section{Learning lifted polynomial models from data}
\label{sec:liftvsextend}
\paragraph{Learning with lifted data.} Since the generators $\g$ are known, the polynomial lifting $\Psi$ can be constructed as described in the proof of Proposition~\ref{prop:existence}.  In order to learn $\F$, it might be tempting to fit it to the \emph{lifted training data}
\(
(\Psi(\x_i), \w_i),
\)
where $\w_i = D\Psi(\x_i)\, \y_i = \F(\Psi(\x_i)) + D\Psi(\x_i) \beps_i$, $\{\Psi(\x_i)\}_{i=1}^N$ are the \emph{lifted states}, $\{\w_i\}_{i=1}^N$ are the \emph{lifted derivatives}, and $\{\bepst_i:=D\Psi(\x_i) \beps_i\}_{i=1}^N$ are the \emph{lifted noise} terms. Although the lifted derivatives are unbiased, i.e., $\E[\w_i] = \F(\Psi(\x_i))$, the lifted noise $\bepst_i$ loses the favorable statistical properties of $\beps_i$, as the next result shows.
\begin{lemma}[Lifted noise]
    \label{lem:lifted_noise}
    For every $i = 1,\dots,N$, the following holds.
    \begin{enumerate}[label=(\roman*), leftmargin=*, noitemsep, topsep=2pt]
        \item The covariance $\bLambda_i$ of the lifted noise $\bepst_i$ is singular.
        \item If $\sigma_k > 0$ for all $k=1, \ldots, n$, then $\bLambda_i$ depends on $\x_i$.
        \item The first $n$ components of $\bepst_i$ are mutually uncorrelated. Moreover, if $\sigma_k > 0$ for all $k=1, \ldots, n$, then for almost every $\x_i \in \Xc'$, each of the components $n+1,\dots,n+\ngen$ of $\bepst_i$ is correlated with at least one of the first $n$ components. \hspace*{\fill}$\square$
    \end{enumerate}
\end{lemma}
By Lemma~\ref{lem:lifted_noise}, the components of $\w_i$ are correlated with a state-dependent covariance $\bLambda_i$, which fitting each component independently ignores. Generalized least squares \citep{aitken1936iv,kariya2004generalized} would account for this but requires inverting $\bLambda_i$, which is singular due to the state-dependent linear dependence $\w_{i,n+1:\ntil} = D\bpsi(\x_i)\,\w_{i,1:n}$ among the outputs. Such dependent outputs must be removed before fitting, since otherwise the parameter estimates are meaningless \citep{box1973some}. Moreover, components of $\F$ fitted independently to the lifted data need not satisfy the relation $\Fh_{n+1:\ntil}(\Psi(\x)) = D\bpsi(\x)\,\Fh_{1:n}(\Psi(\x))$ implied by the chain rule, so that trajectories of the learned model initialized on $\Zc$ may leave it.

\paragraph{Learning with extended states.} Rather than learning $\F$ from the lifted data, we learn only its first $n$ components, which coincide with $\P$, and retrieve the remaining components via the exact relation $\F_{n+1:\ntil} = \Q\,\P$. Learning $\P$ requires only the \emph{extended state samples} $\z_i=(\x_i,\g(\x_i))$, $i = 1, \dots, N$, as well as $\z_j^{\val}=(\x_j^{\val},\g(\x_j^{\val}))$, $j = 1, \dots, \Nv$, together with the measured derivatives $\y_i$ and $\y_j^{\val}$, which carry the i.i.d.\ noise $\beps_i$ and $\beps_j^{\val}$, respectively, so that none of the above issues arise. We assume a known upper bound on the degree of $\P$.
\begin{assumption}
    \label{assump:upper_bound_deg}
    There exists a known $D \in \NN$ such that the true degree $\dks$ \eqref{eq:dk_star} satisfies $\dks \leq D$ for all $k = 1, \dots, n$. \hfill $\square$
\end{assumption}
We now choose the candidate functions $\eta_1(\z), \dots, \eta_{M_D}(\z)$ as all monomials in $\z$ of degree at most $D$, and we denote the resulting library by $\bTheta_D(\z) = [\eta_1(\z), \dots, \eta_{M_D}(\z)]$. When the components of $\g$ satisfy algebraic identities, such as $\sin^2 x_{(l)} + \cos^2 x_{(l)} = 1$, some of these monomials are linearly dependent, in which case, with a slight abuse of notation, $\bTheta_D$ denotes the reduced library obtained by removing the redundant monomials, and $M_D \leq \binom{\nz + D}{D}$ denotes the number of remaining ones. By Assumption~\ref{assump:upper_bound_deg}, every $P_k$ lies in $\spn\{\eta_1, \dots, \eta_{M_D}\}$, so that the library is rich enough and regularized sparse regression is applied as follows. Let $\Z = \begin{bmatrix} \z_1 & \cdots & \z_N \end{bmatrix}^\top$, so that the library evaluated at the extended state samples is the matrix $\bTheta_D(\Z) \in \R^{N \times M_D}$ whose $i$-th row is $\bTheta_D(\z_i)$. Then each coefficient vector $\bxi_k \in \R^{M_D}$, $k = 1,\dots,n$, is computed by STRidge \citep{rudy2017data}, which alternates between solving the ridge problem
\begin{equation*}
    \min_{\bxi} \|\mathbf{Y}_{:,k} - \bTheta_D(\Z)\,\bxi\|^2 + \beta\|\bxi\|^2, \quad \beta > 0, \quad \mathbf{Y} = \begin{bmatrix}\y_1 & \cdots & \y_N\end{bmatrix}^\top,
\end{equation*}
on the remaining columns of $\bTheta_D(\Z)$ and setting coefficients below a threshold $\tau > 0$ to zero, until the number of remaining coefficients stabilizes. Since the initial ridge solution on the full library depends only on $\beta$, it is computed once for every $\beta$ and reused for all thresholds $\tau$. Among all pairs $(\beta, \tau)$ in a finite grid $\Gc$, we select the one minimizing the Akaike information criterion (AIC) on the validation data \citep{akaike1974new,mangan2017model},
\begin{equation*}
    \mathrm{AIC}_k = \Nv \ln\!\Big(\frac{\mathrm{RSS}_k}{\Nv}\Big) + 2\,p_k, \quad \mathrm{RSS}_k = \sum_{j=1}^{\Nv} \big(y^{\val}_{j,k} - \bTheta_D(\z_j^{\val})\,\bxi_k\big)^2, \quad p_k = \big|\{\ell : \xi_{k,\ell} \neq 0\}\big|.
\end{equation*}
We then obtain the learned model
\(
    \Ph(\z) = \big(\hat{P}_1(\z), \dots, \hat{P}_n(\z)\big), 
\)
where $\hat{P}_k(\z) = \bTheta_D(\z)\,\bxi_k$. The lifted state $\Psi(\x)=(\x, \bpsi(\x))$ is needed only afterwards to express the Jacobian $D\bpsi(\x)$ as the polynomial matrix $\Q(\Psi(\x))$ to compute the remaining components of $\F$. Replacing $\Psi(\x)$ by $\zt$ then yields a polynomial model \eqref{eq:tilde_z_dyn} of the lifted dynamics with
\begin{equation}
    \label{eq:F_hat}
    \Fh(\zt) =
    \begin{bmatrix}
        \Ph(\bm{\Pi}\zt) \\ \Q(\zt)\, \Ph(\bm{\Pi}\zt)
    \end{bmatrix}, \quad \bm{\Pi} = \begin{bmatrix} \I_{\nz} & \zero \end{bmatrix}.
\end{equation}
\begin{remark}[Model in the original coordinates]
    A model of $\f$ in the original coordinates is obtained by substituting the components $z_{(n+1)}, \dots, z_{(n+\ngen)}$ of $\z$ in $\Ph$ by the generators $g_{1}(\x), \dots, g_{\ngen}(\x)$, respectively, which gives $\fh(\x) = \Ph(\x, \g(\x))$. \hfill $\square$
\end{remark}
\begin{proposition}[Exact manifold invariance under the learned model]
    \label{prop:invariance}
    Let Assumption~\ref{assump:DA} hold, let $\Psi$ be the polynomial lifting of Proposition~\ref{prop:existence}, and let $\Fh$ be given by \eqref{eq:F_hat}. Then the solution of $\dot{\zt} = \Fh(\zt)$ with $\zt_0 = \Psi(\x_0)$, for any $\x_0 \in \Xc'$, satisfies
    \(
        \zt(t;\Psi(\x_0)) = \Psi(\xh(t;\x_0))
    \)
    for all $t \in [0, \Thmax)$, where $\xh(t;\x_0)$ is the solution of $\dot{\x} = \fh(\x) = \Ph(\x, \g(\x))$ initialized at $\x_0$ and $\Thmax$ is its maximal existence time in $\Xc'$. Thus, the submanifold $\Zc$ is forward invariant under $\Fh$ on $[0, \Thmax)$. \hfill $\square$
\end{proposition}
\section{Kernel-guided SINDy}
\label{sec:KINDy}
\begin{figure}[t]
    \centering
    \begin{tikzpicture}[
        node distance=0.7cm,
        box/.style={draw, rounded corners=2pt, align=center, font=\small, minimum height=0.95cm, inner sep=3pt},
        cheap/.style={box, fill=blue!8},
        exp/.style={box, fill=red!8},
        arr/.style={-{Stealth[length=2mm]}, thick},
        lab/.style={font=\small\bfseries, anchor=east}]
        \node[box] (d2) {$\D, \Dv$};
        \node[cheap, right=of d2] (k2) {KRR in $\z$};
        \node[exp, right=1.3cm of k2] (s2) {STRidge on $\bTheta_{\dkh}$\\ AIC over $(\beta,\tau)$};
        \node[box, right=of s2] (m2) {$\hat{P}_k(\z) = \bTheta_{\dkh}(\z)\,\bxi_k$};
        \draw[arr] (d2) -- (k2); \draw[arr] (k2) -- node[above,font=\small]{$\dkh$} (s2);
        \draw[arr] (s2) -- (m2);
        \node[lab] at ([xshift=-0.2cm]d2.west) {KINDy};
        \node[box, above=0.35cm of d2] (d1) {$\D, \Dv$};
        \node[exp, above=0.35cm of s2] (s1) {STRidge on $\bTheta_D$\\ AIC over $(\beta,\tau)$};
        \node[box, above=0.35cm of m2] (m1) {$\hat{P}_k(\z) = \bTheta_{D}(\z)\,\bxi_k$};
        \draw[arr] (d1) -- (s1); \draw[arr] (s1) -- (m1);
        \node[lab] at ([xshift=-0.2cm]d1.west) {STRidge};
    \end{tikzpicture}
    \caption{STRidge on the full library $\bTheta_D$ (top) versus KINDy (bottom).}
    \label{fig:kindy}
\end{figure}
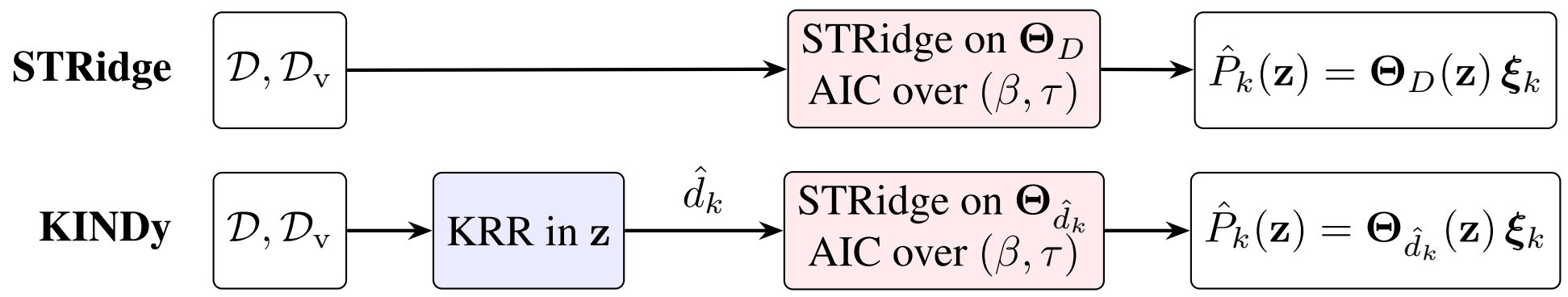
The approach in Section~\ref{sec:liftvsextend} applies STRidge to the polynomial library $\bTheta_D$, whose size satisfies $M_D= \Oc(D^{\nz})$ and therefore increases rapidly with the degree bound $D$. Moreover, STRidge is run for every component $k = 1, \dots, n$ and every pair $(\beta, \tau) \in \Gc$, which multiplies the cost of a single regression, since the refits for small thresholds $\tau$ still act on nearly the entire library $\bTheta_D$. Hence, the resulting cost may become prohibitive for complex systems, since a single bound $D$ must cover the most complex component of $\f$ and is typically chosen conservatively. Ideally, each component $k$ would instead be learned with its own library $\bTheta_{\dks}$, but the true degrees are not known a priori. This motivates finding a degree $\dkh \leq D$ for each component at a cost independent of $M_D$, such that $\dkh$ ideally coincides with $\dks$ and at least does not fall below it. STRidge can then be applied to the smaller libraries $\bTheta_{\dkh}$ without excluding any monomial of the true components.

Kernel ridge regression (KRR)~\citep{scholkopf2001learning} with a polynomial kernel provides such a selection. By the multinomial theorem, the kernel $\kappa_d(\z, \z') = (1 + \z^\top \z')^d$ satisfies $\kappa_d(\z, \z') = \bvarphi_d(\z)^\top \bvarphi_d(\z')$, where the feature map $\bvarphi_d$ consists of all monomials in $\z$ of degree at most $d$, each scaled by the square root of a positive multinomial coefficient. KRR of degree $d$ is therefore a ridge regression on the same monomials as $\bTheta_d$ before the removal of redundant ones, and it can represent $P_k$ exactly whenever $d \geq \dks$. The redundant monomials need not be removed, since KRR only serves to select the degrees $\dkh$ passed on to STRidge, while the regularization keeps the problem well posed. Moreover, KRR can be computed without forming the monomials explicitly, since it requires them only through the Gram matrix $\K_d = [\kappa_d(\z_i, \z_j)]_{i,j=1}^{N} = \bPhi_d \bPhi_d^\top$ with the feature matrix $\bPhi_d = [\bvarphi_d(\z_1) \, \cdots \, \bvarphi_d(\z_N)]^\top$, whose entries are evaluated directly by the kernel. For each degree $d \in \{1, \dots, D\}$, each regularization parameter $\lambda$ from a finite grid $\Omega$, and each component $k = 1, \dots, n$, KRR is fitted to the extended state samples $\z_i$, which gives
\begin{equation}
    \label{eq:krr_degree}
    \hat{P}_k^{d,\lambda}(\z) = \kappa_d(\Z, \z)^\top \balpha_k^{d,\lambda}, \qquad \balpha_k^{d,\lambda} = (\K_d + \lambda \I)^{-1} \Y_{:,k}.
\end{equation}
The prediction \eqref{eq:krr_degree} is the dual form of ridge regression on the features $\bvarphi_d$, which follows from the identity $(\bPhi_d^\top \bPhi_d + \lambda \I)^{-1} \bPhi_d^\top = \bPhi_d^\top (\bPhi_d \bPhi_d^\top + \lambda \I)^{-1}$. Its cost $\Oc(N^2 \nz + N^3)$ does not depend on the number of monomials, so that KRR can be fitted for every degree at a cost independent of $M_D$, while for small $d$ the cheaper primal form, i.e., the ridge problem with $\bPhi_d^\top \bPhi_d$, is used instead. The degree of each component is then selected by minimizing the validation error over both the degree and the regularization parameter, i.e.,
\begin{equation}
    \label{eq:d_hat}
    \dkh = \argmin_{d \in \{1, \dots, D\}} \; \min_{\lambda \in \Omega} V_k(d, \lambda), \qquad V_k(d, \lambda) = \sum_{j=1}^{\Nv} \big(y_{j,k}^{\val} - \hat{P}_k^{d,\lambda}(\z_j^{\val})\big)^2.
\end{equation}
The next result shows that, under a rank condition on the data, $\dkh$ does not fall below the true degree when the noise and the regularization are sufficiently small.
\begin{theorem}[No underestimation of $\dks$]
    \label{thm:no_underestimation}
    Let Assumption~\ref{assump:upper_bound_deg} hold, let $d_k^\star \geq 2$, let $\bPhi^{\val}_d$ be the feature matrix at the validation points $\z_j^{\val}$, let $r_d$ be the maximal number of monomials of degree at most $d$ in $\z$ that are linearly independent on $\Xc$, and let $\f^{\val}_k = [f_k(\x_1^{\val}) \; \ldots \; f_k(\x_{\Nv}^{\val})]^\top$. Suppose that
    \(
        \rank
        \begin{bmatrix}
            \bPhi^{\val}_{\dks-1} &  \f^{\val}_k
        \end{bmatrix} = r_{\dks-1}+1 \text{ and }
        \rank\bPhi_{\dks}=r_{\dks}.
    \)
    Then there exist $\epsb>0$ and $\lamb>0$ such that $\dkh\ge \dks$ whenever $|\epsilon_{i,k}|\le\epsb$ and $|\epsilon^{\val}_{j,k}|\le\epsb$ for all training and validation data, provided that $\Omega$ contains at least one $\lambda\in(0,\lamb]$. \hfill {$\square$}
\end{theorem}
Once the degrees $\dkh$ are computed, each coefficient vector $\bxi_k \in \R^{M_{\dkh}}$ is computed by STRidge on $\bTheta_{\dkh}(\Z)$ and $\Y_{:,k}$ with the model selection by AIC as in Section~\ref{sec:liftvsextend}, but with $D$ replaced by $\dkh$. While Theorem~\ref{thm:no_underestimation} does not exclude overestimation, i.e., $\dkh > \dks$, this only enlarges the library $\bTheta_{\dkh}$ and thus increases the computational cost without excluding any monomial of the true component $P_k$. The learned model then reads $\hat{P}_k(\z) = \bTheta_{\dkh}(\z)\,\bxi_k$, and since $\Ph$ is again polynomial in $\z$, the lifted model \eqref{eq:F_hat} and Proposition~\ref{prop:invariance} hold. We call this method kernel-guided sparse identification of nonlinear dynamics, or KINDy for short, which is illustrated in Figure~\ref{fig:kindy}. Since KRR only selects the degrees, KINDy is not limited to STRidge and can be combined with any sparse regression method, such as the lasso \citep{tibshirani1996regression}.
\paragraph{Computational cost of KINDy.}
The total cost of KINDy consists of the KRR cost for degree selection and the cost of STRidge on the reduced libraries. The cost of a ridge regression on $M$ monomials in the cheaper of the primal and dual forms is $C_{\mathrm{STR}}(M) = \Oc(\min\{N M^2 + M^3, N^2 M + N^3\})$ when the monomials are formed explicitly, as in STRidge, and $C_{\mathrm{KRR}}(M) = \Oc(\min\{N M^2 + M^3, N^2 \nz + N^3\})$ when the dual form evaluates the kernel directly, as in KRR. Since STRidge requires at most $|\Gc|$ such regressions per component, up to a constant number of refits, the leading cost of KINDy is at most $\sum_{d=1}^{D} C_{\mathrm{KRR}}\big(\tbinom{\nz+d}{d}\big) + |\Gc| \sum_{k=1}^{n} C_{\mathrm{STR}}(M_{\dkh})$. The first term, which is the KRR cost with $\tbinom{\nz+d}{d}$ features of degree $d$, holds since $\K_d$, or $\bPhi_d^\top \bPhi_d$ in the primal form, depends neither on $\lambda$ nor on $k$, so that a single eigendecomposition of this matrix per degree yields \eqref{eq:krr_degree} for all $\lambda \in \Omega$ and all components. In comparison, the leading cost of STRidge on $\bTheta_D$ is at most $|\Gc|\, n\, C_{\mathrm{STR}}(M_D)$, and sharing the initial ridge solution across all thresholds lowers the actual cost of both methods below these bounds. KINDy is therefore advantageous whenever $N$ is moderate and the libraries $\bTheta_{\dkh}$ are considerably smaller than $\bTheta_D$, which is typically the case for complex systems whose components differ in complexity.
\section{Example: Cart-pole system}
\label{sec:cart_pole}
\paragraph{Setup.} We consider the unactuated cart-pole system sketched in Figure~\ref{fig:cart_pole} (Appendix \ref{app:supp_cart}). The state is $\x = (p, \theta, \dot{p}, \dot{\theta})$, where $p$ denotes the cart position, and $\theta$ the pole angle. The true model, as well as all system parameters, is also given in Appendix \ref{app:supp_cart}. Since the first two components of the model hold by definition, only the accelerations $\ddot{p}$ and $\ddot{\theta}$ are learned. The data are generated by evaluating the true model at states sampled uniformly from a bounded set and adding zero-mean Gaussian noise to the accelerations. Further details on the experimental setup are given in Table~\ref{tab:settings}.

We assume knowledge of the generators $\g$ satisfying Assumption~\ref{assump:DA} from the underlying physics. More precisely, the revolute joint gives rise to trigonometric functions ($\sin\theta$ and $\cos\theta$), and the rational term $w(\theta) = 1/(1+\bar{\mu}\sin^2\theta)$ is a consequence of the rigid multibody dynamics. Thus, $\g(\x) = (\sin\theta, \cos\theta, w(\theta))$, which is parameter-dependent as it requires the knowledge of $\bar{\mu}$. The extended state is then
\(
    \z = (p, \theta, \dot{p}, \dot{\theta}, \sin\theta, \cos\theta, w(\theta))
\) with $\nz = 7$.
Hence, the two accelerations in Figure~\ref{fig:cart_pole} are given by the following polynomial maps
\(
    P_3(\z) = \bar{\mu} l\, z_{(4)}^2 z_{(5)} z_{(7)} - \bar{\mu}\gamma\, z_{(5)} z_{(6)} z_{(7)}, \, P_4(\z) = \frac{(1+\bar{\mu})\gamma}{l}\, z_{(5)} z_{(7)} - \bar{\mu}\, z_{(4)}^2 z_{(5)} z_{(6)} z_{(7)},
\)
whose true degrees $d_3^\star = 4$ and $d_4^\star = 5$ are assumed to be unknown a priori to the learning methods. Note that the reduced library $\bTheta_D$ is obtained by removing all monomials divisible by $z_{(6)}^2$ or $z_{(5)}^2 z_{(7)}$, which are redundant due to the identities $z_{(5)}^2 + z_{(6)}^2 = 1$ and $z_{(7)}(1 + \bar{\mu} z_{(5)}^2) = 1$.
\paragraph{Results.}
\begin{figure}[t]
    \centering
    \includegraphics[width=0.35\linewidth]{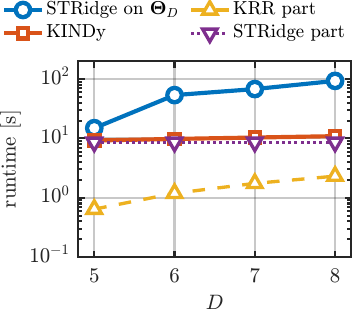}\hspace{2em}
    \includegraphics[width=0.35\linewidth]{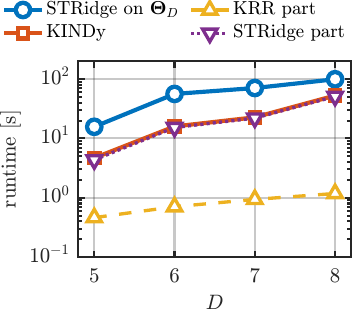}
    \vspace{-4pt}
    \caption{Median runtime versus the bound $D$ at noise levels $1\%$ (left) and $5\%$ (right).}
    \label{fig:runtime}
\end{figure}
We evaluate STRidge on the full library $\bTheta_D$ and KINDy for the bounds $D \in \{5,6,7,8\}$ in terms of the selected degrees, the accuracy, and the runtime. Both methods use model selection by AIC as well as the same data, STRidge implementation, and grid $\Gc$.

\noindent \emph{Degree estimation.} For the low noise levels $1\%$ and $2\%$, KRR returns the true degrees $\hat{d}_3 = d_3^\star = 4$ and $\hat{d}_4 = d_4^\star = 5$ in all seeds and for all bounds $D$, except for one seed at $2\%$ and $D = 8$. For the high noise level $5\%$, KRR returns $\hat{d}_3 = d_3^\star$ in all seeds and for all bounds $D$, whereas $\hat{d}_4$ equals the bound $D$ instead of $d_4^\star$ in 9, 10, and 10 of the 10 seeds for $D = 6$, $7$, and $8$, respectively, as also reflected in the mean predicted degrees in Table~\ref{tab:full_accuracy}. Degree underestimation does not occur for any seed, bound $D$, or noise level, which is consistent with Theorem~\ref{thm:no_underestimation}.

\noindent \emph{Accuracy.} We compare the accuracy of the learned models by the normalized root-mean-square error $e_k = \big(\frac{1}{\Nt}\sum_{i=1}^{\Nt} \big(\hat{P}_k(\z_i^{\test}) - P_k(\z_i^{\test})\big)^2\big)^{1/2} / \sigma_k^{\test}$ on the noise-free extended test states $\z_i^{\test}$, $i = 1, \dots, \Nt$, where $\sigma_k^{\test}$ is the standard deviation of the true values $P_k(\z_i^{\test})$. The median of $e_k$ over the seeds and the mean number of selected monomials are reported in Table~\ref{tab:accuracy} for $D \in \{5, 8\}$ and in Table~\ref{tab:full_accuracy} for all bounds. For the low noise level $1\%$, STRidge on $\bTheta_D$ selects about 4.5 monomials for $P_3$ already for $D = 5$, and this number grows to about nine for $D = 8$, although $P_3$ consists of only two monomials. In all seeds, KINDy predicts $\hat{d}_3 = d_3^\star$ and selects only two monomials for all $D$. Moreover, its median test error is about 0.6 to 0.7 times that of STRidge on $\bTheta_D$. For $P_4$, both methods select two monomials for all $D$ with identical test errors, and KINDy coincides with STRidge on $\bTheta_D$ for $D = d_4^\star = 5$, since both then use the same library. The results for the low noise level $2\%$ are similar. For the high noise level $5\%$, both methods select spurious monomials, but KINDy selects fewer monomials for $P_3$ and attains a lower or equal test error in all cases except for $P_3$ at $D = 8$.

\noindent \emph{Speedup.} For the low noise level $1\%$, the STRidge part of KINDy is the same for all bounds $D$, since KRR always returns the true degrees, while the KRR part increases only slightly with $D$. KINDy thus achieves speedups of 1.6, 5.6, 6.7, and 8.5 over STRidge on $\bTheta_D$ for $D = 5$, $6$, $7$, and $8$, respectively, where the speedup is smallest for $D = 5$, since $D$ then equals $d_4^\star$, so that only the library for $P_3$ is reduced. For the low noise level $2\%$, the speedups are smaller, reaching at most 5.6. For the high noise level $5\%$, the STRidge part of KINDy increases with $D$, since $\hat{d}_4 = D$ in almost all seeds for $D \geq 6$, but KINDy still achieves speedups of 3.4, 3.5, 3.1, and 1.9 for the same bounds. Figure~\ref{fig:runtime} shows the runtimes at the noise levels $1\%$ and $5\%$.  

\noindent \emph{The lifted model.} For the first seed at $1\%$ noise and $D = 6$, KINDy learns
\[
    \hat{P}_3(\z) = 0.25\, z_{(4)}^2 z_{(5)} z_{(7)} - 4.879\, z_{(5)} z_{(6)} z_{(7)}, \quad
    \hat{P}_4(\z) = 29.44\, z_{(5)} z_{(7)} - 0.4997\, z_{(4)}^2 z_{(5)} z_{(6)} z_{(7)},
\]
which is almost identical to the true model. STRidge on $\bTheta_D$, on the other hand, learns the same $\hat{P}_4$ but five monomials in $\hat{P}_3$. Since $\g$ is closed under differentiation, $\bpsi = \g$ and $\ntil = \nz$, so that $\Fh_{5:7}(\zt) = \Q(\zt)\,\Ph(\zt)$ with $\hat{P}_1(\z) = z_{(3)}$, $\hat{P}_2(\z) = z_{(4)}$, where the only nonzero column of $\Q$ is $\Q_{:,2}(\zt) = (\tilde{z}_{(6)}, -\tilde{z}_{(5)}, -\tilde{z}_{(5)}\tilde{z}_{(6)}\tilde{z}_{(7)}^2)$ for $\bar{\mu} = 0.5$. Using $\Fh$ for SOS-based analysis, e.g., to construct barrier functions that bound the cart position $p$, is left for future work.
\begin{table}[t]
	\centering
	\footnotesize
	\caption{Median test error $e_k$, mean predicted degree, and mean number of selected monomials for $D \in \{5, 8\}$ over all seeds. Further results are given in Table \ref{tab:full_accuracy} in Appendix \ref{app:supp_cart}.}
	\label{tab:accuracy}
	\begin{tabular}{lllcccccc}
		\toprule
		& & & \multicolumn{2}{c}{test error ($\times 10^{-3}$)} & \multicolumn{2}{c}{predicted degree} & \multicolumn{2}{c}{selected monomials} \\
		\cmidrule(lr){4-5} \cmidrule(lr){6-7} \cmidrule(lr){8-9}
		noise & $D$ & method & $P_3$ & $P_4$ & $\hat{d}_3$ & $\hat{d}_4$ & $P_3$ & $P_4$ \\
		\midrule
		&   & true  & -- & -- & 4 & 5 & 2 & 2 \\
		\midrule
		$1\%$ & 5 & STRidge on $\bTheta_D$ & 0.64 & 0.35 & -- & -- & 4.5 & 2.0 \\
		&   & KINDy & 0.43 & 0.35 & 4.0 & 5.0 & 2.0 & 2.0 \\
		& 8 & STRidge on $\bTheta_D$ & 0.75 & 0.35 & -- & -- & 8.7 & 2.0 \\
		&   & KINDy & 0.43 & 0.35 & 4.0 & 5.0 & 2.0 & 2.0 \\
		\midrule
		$5\%$ & 5 & STRidge on $\bTheta_D$ & 5.37 & 2.53 & -- & -- & 6.5 & 2.4 \\
		&   & KINDy & 4.49 & 2.53 & 4.0 & 5.0 & 4.5 & 2.4 \\
		& 8 & STRidge on $\bTheta_D$ & 3.84 & 3.26 & -- & -- & 14.6 & 3.8 \\
		&   & KINDy & 4.49 & 3.26 & 4.0 & 8.0 & 4.5 & 3.8 \\
		\bottomrule
	\end{tabular}
\end{table}
\section{Conclusion}
\label{sec:conclusion}
This paper addressed learning polynomial models in lifted coordinates from data when the generators are known from the underlying physics. Since fitting to the lifted data is problematic, we proposed to learn in the extended state and to obtain the remaining dynamics via the exact lifting equations. To reduce the cost of sparse regression on large libraries, we proposed KINDy, which selects the degree of each component by kernel ridge regression before applying STRidge to the correspondingly reduced library, and proved that the selected degree never underestimates the true degree under a rank condition for sufficiently small noise and regularization. On the cart-pole, KINDy recovered the true degrees at low noise levels and achieved speedups of up to 8.5 over STRidge on the full library, while yielding at least as sparse models with lower or equal test errors in almost all cases. Future work includes generators with unknown parameters, systems with inputs, as well as SOS-based analysis and controller synthesis using the learned lifted models.
\paragraph{Generative AI Use.}
AI tools provided by Anthropic (Claude) were used to assist with language editing and code debugging. The authors reviewed and revised all AI-assisted content and take full responsibility for the manuscript and its results.

\paragraph{Acknowledgments}
{The authors acknowledge support by the Deutsche Forschungsgemeinschaft -- DFG  (project number 519323897).}
\clearpage
\bibliography{references}
\clearpage
\appendix
\section{Proofs}
\label{app:proofs}
\subsection{Proof of Theorem \ref{thm:no_underestimation}}
\begin{proof}
For clarity, we first give a sketch of the proof. We begin by establishing a lower bound on the validation error that holds for every degree $d_k < \dks$. We then establish an upper bound on the validation error at $\dks$, which depends on the training noise $\beps_{(k)} := \big[\epsilon_{1,k} \ \cdots \ \epsilon_{N,k}\big]^\top$, the validation noise $\beps_{(k)}^{\val}:= \big[\epsilon_{1,k}^\mathrm{v} \ \cdots \ \epsilon_{N_\mathrm{v},k}^\mathrm{v}\big]^\top$, and the regularization parameter $\lambda$. Finally, we compare the two bounds and show that, when $\beps_{(k)}$, $\beps_{(k)}^{\val}$, and $\lambda$ are sufficiently small, the upper bound at $\dks$ becomes smaller than the lower bound at every $d_k < \dks$. Thus, under these conditions, the minimizer \eqref{eq:d_hat} yields $\dkh \geq \dks$. Note that for $d_k^\star = 1$, the claim holds trivially, since $\hat{d}_k \geq 1$ by~\eqref{eq:d_hat}.

Throughout the proof, let $\ch(d,\lambda) := \bPhi_d^\top \balpha_k^{d,\lambda}$, where $\bPhi_d$ is the feature matrix of $\kappa_d$ at the training points, so that $\K_d = \bPhi_d \bPhi_d^\top$ and the validation predictions of \eqref{eq:krr_degree} are given by $\bPhi^{\val}_d \ch(d,\lambda)$.

\noindent \textbf{Step 1 (Lower bound on $V_k$ for all $d_k < \dks$).} By the first condition in Theorem \ref{thm:no_underestimation}, $\f_k^{\val} \notin \range \bPhi^{\val}_{\dks - 1}$, hence
\begin{equation*}
    \dist\big(\f_k^{\val}, \range \bPhi^{\val}_{\dks - 1}\big) := \min_{\cc} \big\| \bPhi^{\val}_{\dks - 1} \cc - \f_k^{\val} \big\| > 0 .
\end{equation*}
For every degree $d$, let $a_k(d) := \dist\big(\f_k^{\val}, \range \bPhi^{\val}_d\big)^2$, hence $a_k(\dks - 1) > 0$. Since
\begin{equation*}
    \range \bPhi^{\val}_1 \subseteq \range \bPhi^{\val}_2 \subseteq \cdots \subseteq \range \bPhi^{\val}_{\dks - 1},
\end{equation*}
it follows that $a_k(1) \geq a_k(2) \geq \cdots \geq a_k(\dks - 1)$. Hence, for every $d_k < \dks$ and every $\lambda \in \Omega$, the validation prediction $\bPhi^{\val}_{d_k} \ch(d_k, \lambda)$ satisfies
\begin{equation*}
    \big\| \bPhi^{\val}_{d_k} \ch(d_k, \lambda) - \f_k^{\val} \big\|^2 \geq a_k(d_k) \geq a_k(\dks - 1).
\end{equation*}
Let $\y^{\val}_{(k)} = \big[y^{\val}_{1,k} \ \cdots \ y^{\val}_{\Nv,k}\big]^\top$, then $\y^{\val}_{(k)} = \f_k^{\val} + \beps^{\val}_{(k)}$, and by the triangle inequality, we obtain
\begin{equation*}
    \sqrt{V_k(d_k, \lambda)} \geq \big\| \bPhi^{\val}_{d_k} \ch(d_k, \lambda) - \f_k^{\val} \big\| - \big\| \beps^{\val}_{(k)} \big\| \geq \sqrt{a_k(\dks - 1)} - \sqrt{\Nv}\, \epsb,
\end{equation*}
where the last bound is obtained, since $|\epsilon^{\val}_{j,k}| \leq \epsb$ for all $j$ yields $\big\| \beps^{\val}_{(k)} \big\| \leq \sqrt{\Nv}\, \epsb$.
\smallskip

\noindent \textbf{Step 2 (Upper bound on $V_k$ at $\dks$).} Let 
\(
    \f_k = \big[f_k(\x_1) \ \cdots \ f_k(\x_N)\big]^\top
\)
and $\y_{(k)} := \Y_{:,k} = \f_k + \beps_{(k)}$. It follows from the second condition of Theorem~\ref{thm:no_underestimation} and the definition of $r_{d_k^\star}$ that
\begin{equation*}
  \rank
  \begin{bmatrix}
    \bPhi_{\dks} \\
    \bPhi^{\val}_{\dks}
  \end{bmatrix}
  = r_{\dks}
  = \rank \bPhi_{\dks}.
\end{equation*}
Then there exists a constant matrix $\W$ such that $\bPhi^{\val}_{\dks} = \W \bPhi_{\dks}$. Moreover, since $P_k$ is a polynomial of degree $\dks$, then there exists $\cc_k^\star$ such that
$\f_k = \bPhi_{\dks} \cc_k^\star$ and
$\f_k^{\val} = \bPhi^{\val}_{\dks} \cc_k^\star$. Hence, it follows that $\f_k^{\val} = \W \f_k$. With this and using the triangle inequality,
\begin{equation*}
  \sqrt{V_k(\dks,\lambda)}
  = \big\|\bPhi^{\val}_{\dks}\ch(\dks,\lambda) - \f_k^{\val} - \beps^{\val}_{(k)}\big\|
  \le \big\|\W\big(\bPhi_{\dks}\ch(\dks,\lambda) - \f_k\big)\big\| + \big\|\beps^{\val}_{(k)}\big\|,
\end{equation*}
then we get
\begin{equation*}
  \sqrt{V_k(\dks,\lambda)}
  \le \|\W\|\,\big\|\bPhi_{\dks}\ch(\dks,\lambda) - \f_k\big\| + \sqrt{\Nv}\,\epsb.
\end{equation*}
Replacing $\ch(\dks,\lambda)$ by
$\bPhi_{\dks}^\top \balpha_k^{\dks,\lambda}$,
where $\balpha_k^{\dks,\lambda}$ is given in \eqref{eq:krr_degree},
and using $\K_{\dks} = \bPhi_{\dks} \bPhi_{\dks}^\top$, we get
\begin{equation*}
  \sqrt{V_k(\dks,\lambda)}
  \le \|\W\| \, \big\| \Mc \, \y_{(k)} - \f_k \big\|
  + \sqrt{\Nv} \, \epsb,
\end{equation*}
where
$\Mc := \K_{\dks} \big(\K_{\dks} + \lambda \I\big)^{-1}$. Replacing $\y_{(k)}$ by $\f_k + \beps_{(k)}$
and using the triangle inequality, we get
\begin{equation*}
  \sqrt{V_k(\dks,\lambda)}
  \le \|\W\| \Big( \big\| (\Mc - \I) \f_k \big\|
  + \big\| \Mc \beps_{(k)} \big\| \Big)
  + \sqrt{\Nv} \, \epsb.
\end{equation*}
We now bound the noise term $\| \Mc \beps_{(k)} \|$ and the
regularization term $\| (\Mc - \I) \f_k \|$ separately, and then
combine both bounds into a single upper bound on $V_k(\dks,\lambda)$. Taking the singular value decomposition
$\bPhi_{\dks} = \U \Smat \V^\top$,
with singular values $s_1 \ge s_2 \ge \dots \ge 0$ of which exactly the first
$r_{\dks}$ are positive by the second condition of Theorem \ref{thm:no_underestimation},
$\Mc$ becomes
\begin{equation*}
  \Mc = \U \Dmat \U^\top, \qquad
  \Dmat = \Smat \Smat^\top \big( \Smat \Smat^\top + \lambda \I \big)^{-1},
\end{equation*}
where $\Dmat$ is diagonal with entries $s_i^2 / (s_i^2 + \lambda)$
for $i \le r_{\dks}$ and zeros otherwise. Since the entries of $\Dmat$ lie in $[0,1)$ and $\U$ is orthogonal,
this implies
\begin{equation*}
  \big\| \Mc \beps_{(k)} \big\|
  = \big\| \Dmat \U^\top \beps_{(k)} \big\|
  \le \big\| \U^\top \beps_{(k)} \big\|
  = \big\| \beps_{(k)} \big\|
  \le \sqrt{N} \, \epsb.
\end{equation*}
Similarly, $\Mc - \I = \U (\Dmat - \I) \U^\top$
has eigenvalues $-\lambda / (s_i^2 + \lambda)$ for $i \le r_{\dks}$, which are at most
$\lambda / s_{r_{\dks}}^2$ in absolute value, and $-1$ for $i > r_{\dks}$.
Since $\U$ is orthogonal,
\begin{equation*}
  \big\| (\Mc - \I) \f_k \big\|
  = \big\| (\Dmat - \I) \U^\top \f_k \big\| .
\end{equation*}
Moreover, since $\f_k = \bPhi_{\dks} \cc_k^\star$, the vector
$\U^\top \f_k = \Smat \V^\top \cc_k^\star$ vanishes in all
entries $i > r_{\dks}$, so the eigenvalues $-1$ only multiply zeros, and hence
\begin{equation*}
  \big\| (\Dmat - \I) \U^\top \f_k \big\|
  \le \frac{\lambda}{s_{r_{\dks}}^2} \big\| \U^\top \f_k \big\|
  = \frac{\lambda}{s_{r_{\dks}}^2} \| \f_k \| .
\end{equation*}
Finally, we obtain
\begin{equation}
    \label{eq:upper_bound}
    \sqrt{V_k(\dks,\lambda)}
    \le \frac{\lambda \, \|\W\| \, \|\f_k\|}{s_{r_{\dks}}^2} + \Big( \|\W\| \sqrt{N} + \sqrt{\Nv} \Big) \epsb.
\end{equation}
\textbf{Step 3 (Comparing the bounds).}
As $\epsb \to 0$ and $\lambda \to 0$, the upper bound \eqref{eq:upper_bound} on
$\sqrt{V_k(\dks,\lambda)}$ tends to zero, while the lower bound
$\sqrt{a_k(\dks - 1)} - \sqrt{\Nv} \, \epsb$ on
$\sqrt{V_k(d_k,\lambda)}$ for every $d_k < \dks$, which does not depend on $\lambda$, tends to $\sqrt{a_k(\dks - 1)}>0$. Hence there exist $\epsb > 0$ and $\lamb > 0$ such that the upper bound
is smaller than the lower bound, i.e.,
\begin{equation*}
  \frac{\lamb \, \|\W\| \, \|\f_k\|}{s_{r_{\dks}}^2}
  + \Big( \|\W\| \sqrt{N} + \sqrt{\Nv} \Big) \epsb
  < \sqrt{a_k(\dks - 1)} - \sqrt{\Nv} \, \epsb.
\end{equation*}
This inequality holds for every $\lambda \in (0, \bar{\lambda}]$ and since $\Omega$ contains at least one $\lambda \in (0,\lamb]$, we have
$\min_{\lambda \in \Omega} V_k(\dks,\lambda) < \min_{\lambda \in \Omega} V_k(d_k,\lambda)$
for every $d_k < \dks$. Hence the minimizer in \eqref{eq:d_hat} satisfies $\dkh \ge \dks$.
\end{proof}
\subsection{Proof of Proposition \ref{prop:existence}}
\begin{proof}
    The proof is constructive, and we split it into three steps.
    
    \noindent \textbf{Step 1 (Finitely generated fields and rational representations).} Consider the following differential fields
    \begin{equation*}
        \R\langle \g \rangle = \R\Big(\g, \frac{\partial \g}{\partial x_{(l)}}, \frac{\partial^2 \g}{\partial x_{(l)} \partial x_{(m)}}, \dots\Big), \quad
        \Kc = \R(\x)\langle \g \rangle = \R\Big(\x, \g, \frac{\partial \g}{\partial x_{(l)}}, \frac{\partial^2 \g}{\partial x_{(l)} \partial x_{(m)}}, \dots\Big),
    \end{equation*}
    where $l, m = 1, \dots, n$. Since every component of $\g$ is differential-algebraic by Assumption~\ref{assump:DA}, $\R\langle \g \rangle$ is finitely generated over $\R$ as a nondifferential field \cite[Prop.~B.2]{ohtsuka2005model}, and hence so is $\Kc$, since adjoining $\x$ adds only $n$ generators. We also may write $\Kc = \R(\x, \g, \bm{\zeta})$, where $\bm{\zeta} = (\zeta_1, \dots, \zeta_r)$ collects finitely many functions $\zeta_i \notin \R(\x, \g)$ that are real-analytic on $\Xc$ \cite[Prop.~A.3]{ohtsuka2005model}. Because $\Kc$ is closed under differentiation, every partial derivative of $\g$ and $\bm{\zeta}$ belongs to $\Kc$, so that there exist rational functions $r_l$ (written in lowest terms) satisfying
    \begin{equation*}
        \frac{\partial}{\partial x_{(l)}} \begin{bmatrix} \g \\ \bm{\zeta} \end{bmatrix} = r_l(\x, \g, \bm{\zeta}), \quad l = 1, \dots, n,
    \end{equation*}
    which holds on the open and dense subset $\Xc' \subseteq \Xc$ where none of the denominators $\delta_1, \dots, \delta_s$ of the $r_l$ vanishes \cite[Prop.~A.4]{ohtsuka2005model}.
    
    \noindent \textbf{Step 2 (Augmenting with a reciprocal to obtain $\bpsi$).} Similar to \cite[proof of Thm.~1]{ohtsuka2005model}, but with a single reciprocal, we define the product of all denominators $\delta = \delta_1 \cdots \delta_s$ and add its reciprocal $\mu = 1/\delta$, so that each $1/\delta_k = \mu \prod_{j \neq k} \delta_j$ is a polynomial in $(\x, \g, \bm{\zeta}, \mu)$. We then set $\bpsi = (\g, \bm{\zeta}, \mu)$ and $\Psi = (\x, \bpsi)$. The derivatives of $\g$ and $\bm{\zeta}$ are polynomials in $\Psi$, because every denominator $\delta_k$ in the $r_l$ can now be replaced by multiplication with $\mu \prod_{j \neq k} \delta_j$, and the derivative of $\mu$ is a polynomial in $\Psi$ as well, since
    \begin{equation*}
        \frac{\partial \mu}{\partial x_{(l)}} = -\mu^2\, \frac{\partial \delta}{\partial x_{(l)}},
    \end{equation*}
    where $\partial \delta/\partial x_{(l)}$ is a polynomial in $\Psi$ since $\delta$ is a polynomial in $(\x, \g, \bm{\zeta})$. This gives
    \begin{equation*}
        D\bpsi(\x) = \Q(\Psi(\x)) \text{ for all } \x \in \Xc', \quad \Q \text{ a polynomial matrix}.
    \end{equation*}
    
    \noindent \textbf{Step 3 (Polynomial dynamics).} Since $\f(\x) = \P(\x, \g(\x)) = \P(\bm{\Pi}\Psi(\x))$ with $\P$ as defined in Section~\ref{sec:problem} and $\bm{\Pi}$ as in \eqref{eq:F_hat}, the lifted dynamics read
    \begin{equation*}
        \dot{\Psi} = \F(\Psi) = \begin{bmatrix} \I_n \\ \Q(\Psi) \end{bmatrix} \P(\bm{\Pi}\Psi)
    \end{equation*}
    and are polynomial, and therefore $\Psi$ is a polynomial lifting of \eqref{eq:sys}.
\end{proof}
\subsection{Proof of Lemma \ref{lem:lifted_noise}}
\begin{proof}
    The covariance of $\bepst_i$ is $\bLambda_i = D\Psi(\x_i)\,\bSigma\,D\Psi(\x_i)^\top$. Since $\Psi(\x) = (\x, \bpsi(\x))$, we have
    \begin{equation*}
        D\Psi(\x_i) = \begin{bmatrix} \I_n \\ D\bpsi(\x_i) \end{bmatrix} \in \R^{\ntil \times n}, \qquad
        \bLambda_i = \begin{bmatrix} \bSigma & \bSigma\,D\bpsi(\x_i)^\top \\ D\bpsi(\x_i)\,\bSigma & D\bpsi(\x_i)\,\bSigma\,D\bpsi(\x_i)^\top \end{bmatrix}.
    \end{equation*}
    
    \noindent \emph{(i) Singularity.} Since $\rank\bLambda_i \le \rank D\Psi(\x_i) = n < \ntil$, the covariance $\bLambda_i$ is singular.
    
    \noindent \emph{(ii) Dependence on $\x_i$.} Since the first $\ngen$ components of $\bpsi$ are the generators $\g$, which are non-polynomial, $D\bpsi$ is non-constant, so that at least one column $\partial\bpsi/\partial x_{(k)}$ is non-constant. The corresponding column of the block $D\bpsi(\x_i)\,\bSigma$ is $\sigma_k^2\,\partial\bpsi/\partial x_{(k)}(\x_i)$, which is non-constant since $\sigma_k > 0$, so that $\bLambda_i$ depends on $\x_i$. More precisely, it suffices that $\sigma_k > 0$ for some $k$ such that $\partial\bpsi/\partial x_{(k)}$ is non-constant. Moreover, while the first $n$ components of $\bepst_i$ have the constant variances $\sigma_1^2, \ldots, \sigma_n^2$, the variances $D\psi_j(\x_i)\,\bSigma\,D\psi_j(\x_i)^\top$ of the components $n+1,\dots,\ntil$ typically vary with $\x_i$, so that these components are typically heteroscedastic.
    
    \noindent \emph{(iii) Correlation.} The first $n$ components of $\bepst_i$ have the diagonal covariance $\bSigma$ and are thus mutually uncorrelated. Since $g_j$ is non-constant on the connected set $\Xc$, some entry $\partial g_j/\partial x_{(k)}$ of $Dg_j$ does not vanish identically, and since it is real-analytic on $\Xc$, its zero set has zero measure \citep{mityagin2020zero}. Hence, since $\sigma_k > 0$, the $(n+j,k)$-th entry $\sigma_k^2\,\partial g_j/\partial x_{(k)}(\x_i)$ of $\bLambda_i$ is nonzero for almost every $\x_i \in \Xc'$, therefore the $(n+j)$-th component of $\bepst_i$ is correlated with its $k$-th component.
\end{proof}
\subsection{Proof of Proposition \ref{prop:invariance}}
\begin{proof}
    Since $\fh$ is polynomial in $\x$ and $\g$, it is real-analytic on $\Xc$ (hence also locally Lipschitz on $\Xc$). Thus, solutions of $\dot{\x} = \fh(\x)$ are unique for all $\x_0 \in \Xc\supseteq \Xc'$. By Proposition~\ref{prop:existence} and the definition of $\bm{\Pi}$, $\bm{\Pi}\,\Psi(\x) = \z$, and $\Q(\Psi(\x)) = D\bpsi(\x)$ for all $\x \in \Xc'$. Hence,
    \[
        \Fh(\Psi(\x)) =
        \begin{bmatrix}
            \Ph(\x, \g(\x)) \\ D\bpsi(\x)\, \Ph(\x, \g(\x))
        \end{bmatrix}
        =
        \begin{bmatrix}
            \I_n \\ D\bpsi(\x)
        \end{bmatrix} \fh(\x)
        = D\Psi(\x)\, \fh(\x),
        \quad \text{for all } \x \in \Xc'.
    \]
    Therefore, $\Psi$ is a polynomial lifting of $\dot{\x} = \fh(\x)$ in the sense of Definition~\ref{def:lifting} with $\Fh$ in place of $\F$, and the rest of the proof follows by the same argument that led to \eqref{eq:invariance}.
\end{proof}
\clearpage
\section{Supplementary material for the cart-pole}
\label{app:supp_cart}
\begin{figure}[htbp]
    \centering
    \begin{minipage}[c]{0.48\textwidth}
        \centering
        \resizebox{\linewidth}{!}{%
        \begin{tikzpicture}[>=Latex]
            \fill[pattern=north east lines] (-3.5,-0.15) rectangle (3.5,0);
            \draw[thick] (-3.5,0) -- (3.5,0);

            \draw[thick, fill=gray!20] (-1,0.3) rectangle (1,1.1);
            \node at (0,0.7) {$m_1$};
            \draw[thick, fill=white] (-0.6,0.15) circle (0.15);
            \draw[thick, fill=white] ( 0.6,0.15) circle (0.15);

            \coordinate (P) at (0,1.1);
            \draw[dashed] (P) -- ++(0,2.8);
            \draw[line width=2pt] (P) -- ++(65:2.5) coordinate (B);
            \path (P) -- (B) node[midway, right=6pt] {$l$};
            \fill (P) circle (0.06);
            \draw[thick, fill=gray!50] (B) circle (0.18);
            \node[right=5pt] at (B) {$m_2$};

            \draw[->] ($(P)+(90:1.2)$) arc (90:65:1.2);
            \node at ($(P)+(77.5:1.55)$) {$\theta$};

            \draw (-3,-0.35) -- (-3,-0.65);
            \draw ( 0,-0.35) -- ( 0,-0.65);
            \draw[->] (-3,-0.5) -- (0,-0.5) node[midway, below] {$p$};
        \end{tikzpicture}%
        }
    \end{minipage}\hfill
    \begin{minipage}[c]{0.5\textwidth}
        \centering
        \begin{equation*}
            \dot{\x} =
            \begin{bmatrix}
                \dot{p} \\[2pt]
                \dot{\theta} \\[2pt]
                \dfrac{\bar{\mu} \sin\theta \,\big(l \dot{\theta}^2 - \gamma \cos\theta\big)}{1 + \bar{\mu} \sin^2\theta} \\[12pt]
                \dfrac{(1 + \bar{\mu})\, \gamma \sin\theta - \bar{\mu}\, l \dot{\theta}^2 \sin\theta \cos\theta}{l\,\big(1 + \bar{\mu} \sin^2\theta\big)}
            \end{bmatrix}
        \end{equation*}
    \end{minipage}
    \caption{Cart-pole system: schematic on the left, model on the right.}
    \label{fig:cart_pole}
\end{figure}
\begin{table}[!htbp]
    \centering
    \small
    \renewcommand{\arraystretch}{1.15}
    \begin{threeparttable}
        \caption{Settings of the cart-pole experiment.}
        \label{tab:settings}
        \begin{tabular}{ll}
            \toprule
            Parameter & Value \\
            \midrule
            \multicolumn{2}{l}{\emph{System}} \\
            Mass ratio $\bar{\mu} = m_2/m_1$ & $0.5$ \\
            Pole length $l$ & $0.5\,\mathrm{m}$ \\
            Gravitational acceleration $\gamma$ & $9.81\,\mathrm{m/s^2}$ \\
            \midrule
            \multicolumn{2}{l}{\emph{Data}} \\
            Sampling range of $p$ & $[-6, 6] \,\mathrm{m}$ \\
            Sampling range of $\theta$ & $[-4\pi, 4\pi] \, \mathrm{rad}$ \\
            Sampling range of $\dot{p}$ & $[-4, 4] \, \mathrm{m/s}$ \\
            Sampling range of $\dot{\theta}$ & $[-14, 14] \, \mathrm{rad/s}$ \\
            Training samples $N$ & $1000$ \\
            Validation samples $\Nv$ & $500$ \\
            Test samples $\Nt$ & $2000$ \\
            Noise levels\tnote{a} & $1\%$, $2\%$, $5\%$ \\
            Number of random seeds & $10$ \\
            \midrule
            \multicolumn{2}{l}{\emph{Methods}} \\
            Polynomial degree bounds $D$ & $5$, $6$, $7$, $8$ \\
            KRR regularization grid $\Omega$ & $34$ log-spaced values in $[10^{-10}, 10^{1}]$ \\
            STRidge regularization $\beta$ & $11$ log-spaced values in $[10^{-10}, 1]$ \\
            STRidge threshold $\tau$ & $13$ log-spaced values in $[10^{-4}, 1]$ \\
            STRidge grid $\Gc$ & all pairs $(\beta, \tau)$ of the above values \\
            \bottomrule
        \end{tabular}
        \begin{tablenotes}
            \item[a] Relative to the standard deviation of each acceleration.
        \end{tablenotes}
    \end{threeparttable}
\end{table}
\begin{table}[!t]
	\centering
	\renewcommand{\arraystretch}{1.2}
	\setlength{\tabcolsep}{6pt}
	\caption{Median test errors over the seeds, together with the mean predicted degrees and the mean numbers of selected monomials.}
	\label{tab:full_accuracy}
	\begin{tabular}{lllcccccc}
		\toprule
		& & & \multicolumn{2}{c}{test error ($\times 10^{-3}$)} & \multicolumn{2}{c}{predicted degree} & \multicolumn{2}{c}{selected monomials} \\
		\cmidrule(lr){4-5} \cmidrule(lr){6-7} \cmidrule(lr){8-9}
		noise & $D$ & method & $P_3$ & $P_4$ & $\hat{d}_3$ & $\hat{d}_4$ & $P_3$ & $P_4$ \\
		\midrule
		&   & true  & -- & -- & 4 & 5 & 2 & 2 \\
		\midrule
		$1\%$ & 5 & STRidge on $\bTheta_D$ & 0.64 & 0.35 & -- & -- & 4.5 & 2.0 \\
		&   & KINDy & 0.43 & 0.35 & 4.0 & 5.0 & 2.0 & 2.0 \\
		& 6 & STRidge on $\bTheta_D$ & 0.65 & 0.35 & -- & -- & 4.6 & 2.0 \\
		&   & KINDy & 0.43 & 0.35 & 4.0 & 5.0 & 2.0 & 2.0 \\
		& 7 & STRidge on $\bTheta_D$ & 0.75 & 0.35 & -- & -- & 6.8 & 2.0 \\
		&   & KINDy & 0.43 & 0.35 & 4.0 & 5.0 & 2.0 & 2.0 \\
		& 8 & STRidge on $\bTheta_D$ & 0.75 & 0.35 & -- & -- & 8.7 & 2.0 \\
		&   & KINDy & 0.43 & 0.35 & 4.0 & 5.0 & 2.0 & 2.0 \\
		\midrule
		$2\%$ & 5 & STRidge on $\bTheta_D$ & 2.01 & 0.70 & -- & -- & 7.0 & 2.0 \\
		&   & KINDy & 0.86 & 0.70 & 4.0 & 5.0 & 2.7 & 2.0 \\
		& 6 & STRidge on $\bTheta_D$ & 1.66 & 0.96 & -- & -- & 6.4 & 2.2 \\
		&   & KINDy & 0.86 & 0.70 & 4.0 & 5.0 & 2.7 & 2.0 \\
		& 7 & STRidge on $\bTheta_D$ & 1.29 & 0.70 & -- & -- & 7.0 & 2.0 \\
		&   & KINDy & 0.86 & 0.70 & 4.0 & 5.0 & 2.7 & 2.0 \\
		& 8 & STRidge on $\bTheta_D$ & 1.44 & 0.70 & -- & -- & 9.9 & 2.0 \\
		&   & KINDy & 0.86 & 0.70 & 4.0 & 5.3 & 2.7 & 2.0 \\
		\midrule
		$5\%$ & 5 & STRidge on $\bTheta_D$ & 5.37 & 2.53 & -- & -- & 6.5 & 2.4 \\
		&   & KINDy & 4.49 & 2.53 & 4.0 & 5.0 & 4.5 & 2.4 \\
		& 6 & STRidge on $\bTheta_D$ & 7.88 & 3.21 & -- & -- & 8.5 & 3.4 \\
		&   & KINDy & 4.49 & 3.21 & 4.0 & 5.9 & 4.5 & 3.2 \\
		& 7 & STRidge on $\bTheta_D$ & 5.06 & 2.33 & -- & -- & 12.8 & 3.7 \\
		&   & KINDy & 4.49 & 2.33 & 4.0 & 7.0 & 4.5 & 3.7 \\
		& 8 & STRidge on $\bTheta_D$ & 3.84 & 3.26 & -- & -- & 14.6 & 3.8 \\
		&   & KINDy & 4.49 & 3.26 & 4.0 & 8.0 & 4.5 & 3.8 \\
		\bottomrule
	\end{tabular}
\end{table}
\end{document}